\documentclass[aps,prl,twocolumn,superscriptaddress,showpacs,preprintnumbers,amsmath,amssymb]{revtex4}
\usepackage{color}
\usepackage{longtable}
\usepackage{graphicx} % Include figure files
\usepackage{dcolumn}  % Align table columns on decimal point

\graphicspath{{ps}}

\begin{document}

%\vspace*{-3\baselineskip}
%\resizebox{!}{3cm}{\includegraphics{belle.eps}}

\preprint{\vbox{ 
                 %\hbox{Belle Preprint 2008-xx}
                 %\hbox{KEK   Preprint 2008-xx}
                 %\hbox{hep-ex/08-xxxxxx}
}}

\title{ \quad\\[0.5cm] Observation of $B^{\pm} \to \psi(2S) \pi^{\pm}$ and search for direct $CP$-violation }  
%%% Paper: %%% Journal:  Physical Review
%%% Contacts: %%% Non-responding authors or those who said NO are commented out
%%% Use \input{author} to insert this material into your latex file.
%%%%% Force institutions to appear in alphabetical order when typeset.
\affiliation{Budker Institute of Nuclear Physics, Novosibirsk}
\affiliation{Chiba University, Chiba}
\affiliation{University of Cincinnati, Cincinnati, Ohio 45221}
%%%\affiliation{T. Ko\'{s}ciuszko Cracow University of Technology, Krakow}
\affiliation{Department of Physics, Fu Jen Catholic University, Taipei}
\affiliation{Justus-Liebig-Universit\"at Gie\ss{}en, Gie\ss{}en}
\affiliation{The Graduate University for Advanced Studies, Hayama}
\affiliation{Gyeongsang National University, Chinju}
\affiliation{Hanyang University, Seoul}
\affiliation{University of Hawaii, Honolulu, Hawaii 96822}
\affiliation{High Energy Accelerator Research Organization (KEK), Tsukuba}
\affiliation{Hiroshima Institute of Technology, Hiroshima}
%%%\affiliation{University of Illinois at Urbana-Champaign, Urbana, Illinois 61801}
\affiliation{Institute of High Energy Physics, Chinese Academy of Sciences, Beijing}
\affiliation{Institute of High Energy Physics, Vienna}
\affiliation{Institute of High Energy Physics, Protvino}
\affiliation{Institute for Theoretical and Experimental Physics, Moscow}
\affiliation{J. Stefan Institute, Ljubljana}
\affiliation{Kanagawa University, Yokohama}
\affiliation{Korea University, Seoul}
%%%\affiliation{Kyoto University, Kyoto}
\affiliation{Kyungpook National University, Taegu}
\affiliation{\'Ecole Polytechnique F\'ed\'erale de Lausanne (EPFL), Lausanne}
\affiliation{Faculty of Mathematics and Physics, University of Ljubljana, Ljubljana}
\affiliation{University of Maribor, Maribor}
\affiliation{University of Melbourne, School of Physics, Victoria 3010}
\affiliation{Nagoya University, Nagoya}
\affiliation{Nara Women's University, Nara}
\affiliation{National Central University, Chung-li}
\affiliation{National United University, Miao Li}
\affiliation{Department of Physics, National Taiwan University, Taipei}
\affiliation{H. Niewodniczanski Institute of Nuclear Physics, Krakow}
\affiliation{Nippon Dental University, Niigata}
\affiliation{Niigata University, Niigata}
\affiliation{University of Nova Gorica, Nova Gorica}
\affiliation{Osaka City University, Osaka}
\affiliation{Osaka University, Osaka}
\affiliation{Panjab University, Chandigarh}
%%%\affiliation{Peking University, Beijing}
%%%\affiliation{Princeton University, Princeton, New Jersey 08544}
%%%\affiliation{RIKEN BNL Research Center, Upton, New York 11973}
\affiliation{Saga University, Saga}
\affiliation{University of Science and Technology of China, Hefei}
\affiliation{Seoul National University, Seoul}
%%%\affiliation{Shinshu University, Nagano}
\affiliation{Sungkyunkwan University, Suwon}
\affiliation{University of Sydney, Sydney, New South Wales}
\affiliation{Tata Institute of Fundamental Research, Mumbai}
\affiliation{Toho University, Funabashi}
\affiliation{Tohoku Gakuin University, Tagajo}
%%%\affiliation{Tohoku University, Sendai}
\affiliation{Department of Physics, University of Tokyo, Tokyo}
\affiliation{Tokyo Institute of Technology, Tokyo}
\affiliation{Tokyo Metropolitan University, Tokyo}
\affiliation{Tokyo University of Agriculture and Technology, Tokyo}
%%%\affiliation{Toyama National College of Maritime Technology, Toyama}
\affiliation{Virginia Polytechnic Institute and State University, Blacksburg, Virginia 24061}
\affiliation{Yonsei University, Seoul}
 \author{V.~Bhardwaj}\affiliation{Panjab University, Chandigarh} % Panjab
\author{R.~Kumar}\affiliation{Panjab University, Chandigarh} % Panjab
 \author{J.~B.~Singh}\affiliation{Panjab University, Chandigarh} % Panjab
 \author{I.~Adachi}\affiliation{High Energy Accelerator Research Organization (KEK), Tsukuba} % KEK
 \author{H.~Aihara}\affiliation{Department of Physics, University of Tokyo, Tokyo} % Tokyo
% \author{D.~Anipko}\affiliation{Budker Institute of Nuclear Physics, Novosibirsk} % BINP
 \author{K.~Arinstein}\affiliation{Budker Institute of Nuclear Physics, Novosibirsk} % BINP
% \author{T.~Aso}\affiliation{Toyama National College of Maritime Technology, Toyama} % Toyama
% \author{V.~Aulchenko}\affiliation{Budker Institute of Nuclear Physics, Novosibirsk} % BINP
 \author{T.~Aushev}\affiliation{\'Ecole Polytechnique F\'ed\'erale de Lausanne (EPFL), Lausanne}\affiliation{Institute for Theoretical and Experimental Physics, Moscow} % ITEP
 \author{T.~Aziz}\affiliation{Tata Institute of Fundamental Research, Mumbai} % Tata
 \author{S.~Bahinipati}\affiliation{University of Cincinnati, Cincinnati, Ohio 45221} % Cincinnati
 \author{A.~M.~Bakich}\affiliation{University of Sydney, Sydney, New South Wales} % Sydney
% ...
% BINP
\collaboration{The Belle Collaboration}

%\collaboration{Belle Collaboration}

\begin{abstract}
We  report the  first observation of $B^{\pm}\to\psi(2S)\pi^{\pm}$, 
a Cabibbo- and color-suppressed decay. This analysis is based 
on $657\times10^{6}$ $B\overline B$ events collected at the 
$\Upsilon(4S)$ resonance with the Belle detector at the KEKB 
 energy-asymmetric $e^+e^-$ collider. The measured branching fraction 
is  ($2.44 \pm 0.22 \pm 0.20$)$\times 10^{-5}$ and  the charge asymmetry is  $\mathcal{A}=0.022 \pm 0.085 \pm 0.016$.  The ratio of the
branching fractions  
$\mathcal {B}( B^{\pm} \to \psi(2S) \pi^{\pm})$$/
\mathcal {B}( B^{\pm} \to \psi(2S) K^{\pm})$ $  = 
(3.99 \pm 0.36\pm 0.17)\%$ is also determined.
\end{abstract}

%Use showkeys class option if keyword

\maketitle

The study of exclusive  $B$ meson decays to charmonium  has played
an important role in  exploring $CP$-violation~\cite{bdecays}. Among these, Cabibbo-suppressed charmonium decays provide an opportunity to understand  nonleptonic $B$ decays \cite{avery,aubert1,abe1,aubert2,jpsipi0,
 babarjpsipi0,rajeev,jeri} and can be  studied at the  $B$ factories which, due to their high integrated  luminosities, overcome the suppression factor. At  quark level, the decay  $B^- \to \psi(2S) \pi^-$ \cite{chargeconju} proceeds primarily via a $b \to c \overline{c} {d}$ transition; the leading contribution to this decay comes from the Cabibbo- and color-suppressed tree diagram shown in  Fig.~1. The measurement  of the  branching fraction for the decay  $B^- \to \psi(2S) \pi^-$, which has not been observed so far, is important for detailed studies of the $b \to c \overline{c}d$ transition.

Assuming tree dominance and factorization, the branching fraction 
${\cal B}(B^{-}\to \psi(2S)\pi^{-})$  is expected to be about  
$5\%$ of that of the Cabibbo-favoured  mode 
$B^{-}\to \psi(2S)K^{-}$  \cite{neubert}. 
Under this assumption, the charge asymmetries  for the  
$ B^{-}\to \psi(2S)\pi^{-} $ and  $ B^{-}\to \psi(2S)K^{-} $ decays are 
expected to be negligibly small in the  standard model.  However, the penguin amplitude in  the $b\to c\overline{c}d$ transition contains a 
complex phase due to one of the Kobayashi-Maskawa matrix 
elements \cite{new2}, $V_{td}$, and therefore, if  penguin  
or new physics contributions are substantial, direct $CP$ violation 
may occur in $B^- \to \psi(2S)  \pi^-$ \cite{gronau, dunitez}.

 In this paper, we report the first observation of the 
$B^{-}\to\psi(2S)\pi^{-}$ decay, along with a measurement of the
branching fractions $\mathcal{B}(B^{-} \to \psi(2S) \pi^{-})$  and  the ratio $\mathcal{B}(B^{-} \to \psi(2S) \pi^{-})/\mathcal{B}(B^{-} \to ~\psi(2S) K^{-})$. 
A search for direct $CP$-violation in $B^{-}\to\psi(2S)\pi^{-}$ decays 
is also presented. These measurements are based on a data sample 
that contains $657\times10^{6}~B\overline{B}$ events, collected 
with the Belle detector \cite{abashian} at the KEKB \cite{kurokawa}  
 energy-asymmetric $e^+e^-$ collider operating at the $\Upsilon(4S)$ resonance.

%%%%%%%%%%%%%%%%%%%%%%%%%%%%%%%%%%%%%%%%%%%%%%%%%%%%%%%%%%%%
\begin{figure}
\includegraphics[width=0.45\textwidth]{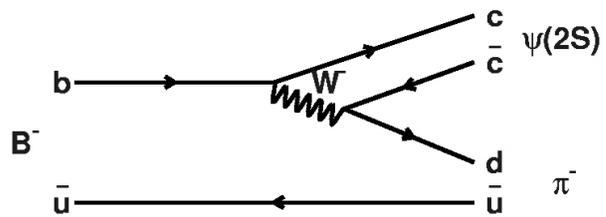}
\caption{Leading-order tree level diagram for the decays under study.}
\end{figure}
%%%%%%%%%%%%%%%%%%%%%%%%%%%%%%%%%%%%%%%%%%%%%%%%%%%%%%%%%%%%

 The Belle detector is a large solid-angle magnetic spectrometer 
that consists of a silicon vertex detector (SVD) surrounded by a 
50-layer central drift chamber (CDC), an array of aerogel 
threshold  Cherenkov counters (ACC), a barrel-like 
arrangement of time-of-flight  (TOF) scintillation counters, and an 
electromagnetic calorimeter (ECL) comprised of CsI(Tl) crystals. 
All these subdetectors are located inside a superconducting solenoid 
coil that provides a $1.5$ T magnetic field. An iron flux-return yoke 
located outside the coil is instrumented to detect ${K^0_L}$ mesons
and to identify the muons (KLM). The data sample used in this analysis is 
collected with two different detector configurations. The first sample 
of  $152 \times 10^6~B\overline{B}$ events is  collected with a
$2.0$~cm radius beam-pipe and a $3$-layer SVD, while  the remaining $505 \times 10^{6}~B\overline{B}$ events  are  collected with a
$1.5$~cm radius  beam-pipe, a $4$-layer SVD and a smaller-cell inner 
drift chamber  \cite{ natkaniec}.   The detector is described in detail 
elsewhere \cite{abashian}. A GEANT-based Monte Carlo (MC) simulation is used to model the response of the detector and determine the efficiency of the 
signal reconstruction \cite{evtgen, geant}.

The reconstruction of the $\psi(2S)$  meson  is performed using the $\ell^+\ell^-$ ($\ell =$ $e$ or $\mu$) and $J/\psi\pi^+ \pi^-$ decay channels. The $J/\psi$ mesons are reconstructed in the $\ell^+ \ell^-$ decay channel.  Both daughter tracks of 
$J/\psi\to\ell^+\ell^-$ or $\psi(2S)\to\ell^+\ell^-$ decays are
required to be positively identified as leptons. From the selected charged tracks, $e^+$ and $e^-$  candidates are identified by combining information from the CDC ($dE/dx$), $E/p$  ($E$ is the energy deposited in the ECL and $p$ is the momentum measured by  the SVD and the CDC),  and shower shape in the ECL.   In addition, ACC information and position matching between   electron track candidates and ECL clusters are used in the identification of electron candidates. Identification of $\mu$ candidates is based on  the track penetration depth  and  hit pattern  in the KLM system.

%%%%%%%%%%%%%%%%%%%%%%%%%%%%%%%%%%%%%%%%%%%%%
  In  $J/\psi$ $\to$ $e^+e^-$ and $\psi(2S)$ $\to$ $e^+e^-$ decays, 
the four-momenta of all photons within 50 mrad of each of the original
$e^+$ or $e^-$ tracks are included in the invariant mass calculation 
$[$hereafter denoted as  $M_{e^+e^- (\gamma)}$$]$, in order to reduce the radiative tail.  The reconstructed  invariant mass of  the $J/\psi$ candidates is required to satisfy $-0.150$ GeV$/c^2 < M_{e^+ e^-(\gamma)} -m_{J/\psi} < 0.036$ GeV$/c^2$  or  $-0.060$ GeV$/c^2  < M_{\mu^+ \mu^-} - m_{J/\psi} < 0.036$ GeV$/c^2$, where   $m_{J/\psi}$ denotes the nominal world-average $J/\psi$ 
mass  \cite{PDG2006}. These intervals are  asymmetric in order to  include 
part of the radiative tails. In $\psi(2S)$ reconstruction, the
invariant mass  is required to satisfy $3.55$ GeV$/c^2$ $<$ 
$M_{e^+  e^- (\gamma )} < 3.75$ GeV$/c^2$ 
 or  $3.65$ GeV$/c^2 < M_{\mu^+ \mu-} < 3.75$ GeV$/c^2$. 
For the $\psi(2S)\to J/\psi \pi^+ \pi^-$ candidates, 
$\Delta M$ $=$ $M_{\ell^+ \ell^- \pi^+ \pi^-}$ $-$ $M_{\ell^+\ell^-}$ 
should satisfy the condition $0.578$ GeV$/c^2$ $ <$ 
$\Delta M$ $<$ $0.598$ GeV$/c^2$.  In order to reduce the
combinatorial background from low-momentum pions, 
 $M_{\pi^+\pi^-}$ (invariant mass of the two pions coming from  the  $\psi(2S)$ decay) is  required  to be 
greater than $0.40$ GeV$/c^2$. In order to veto peaking background 
coming from $K^0_S$ ($B^{-}\to J/\psi {K^*}^{-}(K^0_S \pi^{-})$), 
all pairs of pions  with mass $0.4856$ GeV$/c^2$ $<$ $M_{\pi^+\pi^-}$ $<$ $0.5096$ GeV$/c^2$ are rejected in  the $\psi(2S) \to$ $J/\psi\pi^+ \pi^-$ 
reconstruction  of the $B^{-} \to  \psi(2S) \pi^{-}$ mode.  A  mass- 
and  vertex-constrained fit  is performed to all the 
selected $\psi(2S)$ and $J/\psi$ candidates in order to improve 
the momentum resolution. 
%%%%%%%%%%%%%%%%%%%%%%%%%%%%%%%%%%%%

%%%%%%%%%%%%%%%%%%%%%%%%%%%%%%%%%%%%%%%

 We  combine  $\psi(2S)$ and  $\pi^{-}$ ($K^{-}$) mesons to form  $B$ meson candidates.  The combined information from the CDC $(dE/dx)$, TOF and ACC is used to identify $\pi^{-}$ $(K^{-})$ on the  basis of the $\pi-K$ likelihood ratio, 
$\mathcal{R}_{\pi(K)}$  $=$ 
$ \mathcal{L}_{\pi(K)}/(\mathcal{L}_{\pi} +\mathcal{L}_{K})$, 
where $\mathcal{L}_{\pi}~(\mathcal{L}_{K})$  is the  likelihood of a pion (kaon) hypothesis.  Charged tracks with $\mathcal{R}_{\pi}$ $>$ $0.85$ ($\mathcal{R}_{K}$ $>$ $0.85$)  are identified as  $\pi^-$ ($K^{-}$). This requirement is $88.0\%$ ($82.6\%$) efficient  for $\pi$ ($K$)  with a $K$ ($\pi$) fake rate of $6.0\%$ ($3.7\%$). 

  To discriminate the signal from background, we use  two kinematic variables: the beam-constrained 
mass, $M_{\rm bc}$ $\equiv$ $\sqrt{{{E}^{*2} _{\rm beam}} -{p^{*2}_{B}}}$ 
and  the energy difference, $\Delta E \equiv E_{B}^*- E^*_{\rm beam}$  to discriminate  the 
signal from the background, where ${E^*_{ \rm beam}}$ is the 
run-dependent beam energy, and ${E^*_{B}}$ and ${p^*_{B}}$ 
are the reconstructed energy and momentum, respectively, 
of the $B$ meson candidates in the center-of-mass (CM)  frame. 
We retain $B$ candidates  with $5.27$ GeV$/c^2$ $<$ $M_{\rm bc}$ 
$<$ $5.29$ GeV$/c^2$ and $-0.15$ GeV $<$ $\Delta E$ $<0.2$ GeV.  After all selection requirements, $2.3\%$ of the events for $\psi(2S)(\ell^+\ell^-)\pi^-$ and $7.1\%$ for $\psi(2S)(J/\psi \pi^+ \pi^-) \pi^-$ contain more than one $B$ candidate. For these events, we choose the $B$ candidate whose $M_{\rm bc}$ is closest to the nominal $B$ meson mass. 
%%%%%%%%%%%%%%%%%%%%%%%%%%%%%%%%%%%%%

 To suppress continuum background, events having a ratio of the second to zeroth Fox-Wolfram moments \cite{foxwolfram} $R_2 >0.5$ are rejected.  Large $ B \to ( J/\psi,\psi(2S) )X $  MC samples  corresponding to $3.86 \times 10^{10}$ generic $B\overline{B}$ events 
are used  for the background study, because  backgrounds predominantly 
come from $B$ decays into final states having 
$J/\psi \to \ell^+ \ell^-$ or $\psi(2S) \to \ell^+ \ell^-$. 
We find that the dominant background comes from $B^-\to\psi(2S)K^-$ where the kaon is misidentified as a pion. This decay mode makes a peak at $\Delta E \sim -0.07$ GeV.   Other backgrounds  originate from random combinations of $\psi(2S)$ and $\pi^-$ candidates (combinatorial background) and do not form any peaking structure in the $\Delta E $ projection of the MC sample.  Studies of $\Delta M$ and $\ell^+ \ell^-$ invariant mass data sidebands 
support this assumption.

%%%%%%%%%%%%%%%%
 We extract the signal yield by 
performing  an unbinned extended maximum likelihood fit to 
the   $\Delta E$  distribution of the selected $B$
candidates. The extended likelihood function used is

\begin{equation*}
  \mathcal{L}(N_S, N_{BK}, N_{C})= \frac{e^{-(N_S + N_{BK} + N_C)}}{N!}
 \prod_{i=1}^N   [N_S P_S(\Delta {E}_i)  + \end{equation*}\begin{equation} 
     N_{BK} P_{BK}(\Delta { E}_i) + N_C P_C(\Delta { E}_i) ],   
\end{equation}
where $N$ is the total  number of candidate events; 
$N_S$, $N_{BK}$ and $N_C$ denote the number of signal, background 
from $B^{-}\to\psi(2S)K^{-}$ and  combinatorial  background events, respectively.   The  modeling PDF's include a sum of two Gaussians for the signal ($P_S(\Delta {E}_i)$), a sum of two bifurcated Gaussians for the $B^-\to\psi(2S)K^-$  background ($P_{BK}(\Delta {E}_i)$) and a second-order polynomial for other backgrounds ($P_C(\Delta {E}_i)$).
 In this study the $B^{-}\to$ $\psi(2S)K^{-}$ decay mode is 
used as a control sample, as well as the  denominator of the relative   
branching fraction  ratio ${\cal B}(B^{-}\to \psi(2S)\pi^{-})$$/{\cal
B}(B^{-}\to \psi(2S)K^{-})$.  The $B^- \to \psi(2S) \pi^-$ signal shape  is fixed to that obtained  from the $B^{-}\to \psi(2S) K^{-}$ control sample.  The $B^- \to \psi(2S) K^-$   background shape is determined from the MC sample after applying a correction for the difference between   data and MC. For  the smooth combinatorial background, parameters of the second-order polynomial are floated in the fit.

%%%%%%%%%%%%%%%
\begin{figure}[h!]
\begin{center}
\includegraphics[scale=0.45]{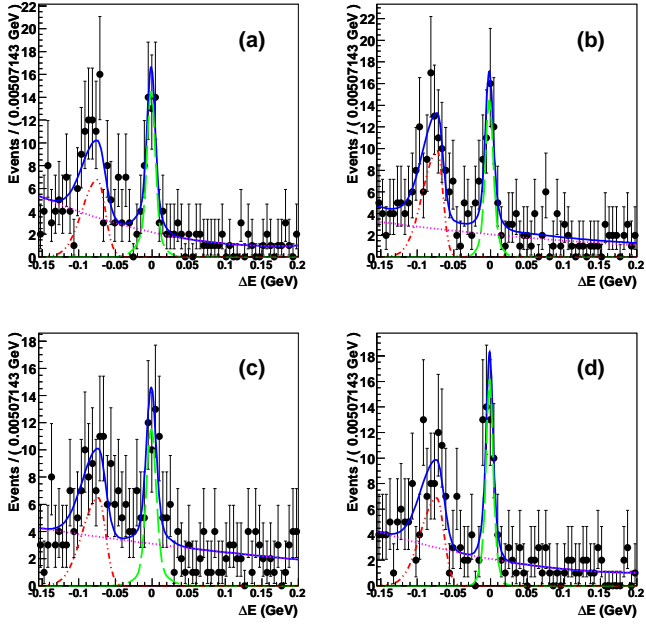}
\caption{\label{fig:qm/complexfunctions} $\Delta E$ distributions of the  $B^{-}\to\psi(2S)\pi^{-}$ candidates, reconstructed in the following four sub-modes: 
(a) $B^{-}\to\psi(2S)(J/\psi(ee)\pi^+\pi^-)\pi^-$, 
(b) $B^{-}\to\psi(2S)(J/\psi(\mu\mu)\pi^+\pi^-)\pi^-$, 
(c) $B^{-}\to\psi(2S)(ee)\pi^-$ and 
(d) $B^{-}\to\psi(2S)(\mu\mu)\pi^-$. The curves show the signal (green dashed) and the background components (red dot-dashed for $B^-\to\psi(2S)K^-$, magenta dotted for  combinatorial background) as well as the overall fit (blue solid). }
\end{center}
\end{figure}
%%%%%%%%%%%%%%%%
 We perform a simultaneous fit to all the considered 
$\psi(2S)$ decay modes to obtain a common branching fraction, 
taking into account different detection efficiencies for each sub-decay mode  and  detector configuration. The efficiencies are determined from
   signal MC samples after applying the correction factors  taking into account  the data and MC differences as described later.   In the fit, the  yield of $B^-\to\psi(2S)K^-$ (background) and the yield and polynomial parameters of the combinatorial backgrounds are floated for each decay mode separately.
   The fit gives a branching fraction of $(2.44 \pm 0.22 \pm 0.20)\times 10^{-5}$, where the first error is statistical and the second is systematic.  Equal production of neutral 
and charged $B$ meson pairs in the $\Upsilon(4S)$ decay is assumed. We use the   $J/\psi\to e^+ e^-$, $J/\psi \to \mu^+ \mu^-$, $\psi(2S)\to J/\psi \pi^+ \pi^-$, $\psi(2S) \to e^+ e^-$ and $\psi(2S) \to \mu^+ \mu^-$  intermediate  branching fractions  from Ref.~\cite{PDG2006}.   Separate fits to the four sub-modes are also performed.  The  results are shown in Fig.~2 and summarized in Table~I. The yields are calculated from the obtained branching fractions and efficiencies.   The efficiencies listed in Table~I are  weighted averages of  the two separate data sets.    For all sub-modes, we observe signals with more than  $5\sigma$ statistical significance   and the branching fractions agree well with  one another.   The statistical significance   is defined  as $\sqrt{-2~{\rm ln}~(\mathcal{L}_0 / \mathcal{L}_{\rm max} )}$ where  $\mathcal{L}_{\rm max}  $ ($\mathcal{L}_{0}$) denotes the likelihood value at the maximum (with the signal yield fixed to zero).

A correction for  small differences in  the signal detection efficiency calculated from signal MC and data has been applied for the pion  identification requirement, muon identification and $\Delta M$. Uncertainties on these corrections  are included in the systematic error. The  $J/\psi \to \ell^+ \ell^-$ and $e^+ e^- \to e^+ e^- \ell^+ \ell^-$ samples are used to 
estimate the lepton identification correction and uncertainty 
whereas  the pion (kaon) identification  correction and 
uncertainty are determined from a $D^{*+} \to D^0(K^-\pi^+) \pi^+$ sample.
A correction factor for the $\Delta M$ requirement is determined 
from the $B^{-}\to\psi(2S)K^{-}$ sample and is estimated by taking 
the ratio of yields from data and MC for  tight ($0.578$ GeV/$c^2<$ $\Delta M$ $<0.598$ GeV/$c^2$) and loose ($0.570$ GeV/$c^2$ $< \Delta M$ $<0.610$ GeV/$c^2$)	 selection cuts.   The fitting procedure is checked using signal and $B\to(J/\psi,\psi(2S)) X$ MC samples and no significant  bias is found.  The signal yield systematic uncertainty is calculated by varying each fixed parameter in the fit by $\pm 1 \sigma$, and then taking the  sum in quadrature of the deviations of the signal yield from the nominal value. The total systematic uncertainty assigned to the yield estimation  is $3.5\%$.  The uncertainties  due to the daughter branching fractions  amount  to  $3.4\%$.  The uncertainty on the tracking efficiency is estimated to be $1.2\%$ per track and  $5.0\%$ in total.  Systematic errors have been estimated by taking into account whether the particular error is correlated or uncorrelated and then combining them with the proper weights.   All the systematic  uncertainties are summarized in Table~II.  The total systematic error of $8.4\%$ is the sum in quadrature of all the  uncertainties. 

{
\begin{table*}
\caption{\label{tab:table1} Summary of the results.  }
\begin{center}
\begin{ruledtabular}
\begin{tabular}{lcccc}
Decay mode & Efficiency(\%) & Signal yield  & Branching fraction($10^{-5}$)  & Statistical  significance \\ \hline
$B^{-}\to\psi(2S)(J/\psi(e^+e^-)\pi^+\pi^-)\pi^{-}$ & 15.1 & 48.9$\pm$8.3 & $2.57\pm 0.44$ & $ 9.5 \sigma$ \\         
$B^{-}\to\psi(2S)(J/\psi(\mu^+\mu^-)\pi^+\pi^-)\pi^{-}$ & 16.8 &  44.0$\pm$8.1 & $2.08\pm0.38$  & $ 8.4\sigma$  \\
$B^{-}\to\psi(2S)(e^+e^-)\pi^{-}$ &  32.2   & 44.0$\pm$9.0 & $2.80\pm0.57$    & $ 7.3 \sigma$       \\
$B^{-}\to\psi(2S)(\mu^+\mu^-)\pi^{-}$ & 35.7  & 43.5$\pm$7.7  & $2.50\pm0.44$  & $ 9.0 \sigma $ \\      \hline    
$B^{-}\to\psi(2S)\pi^{-}$ (combined)  &  &  & $2.44\pm0.22\pm0.20$ &  \\ 
\end{tabular}
\end{ruledtabular}
\end{center}
\end{table*}

}

\begin{table}
\caption{\label{tab:table2} Summary of systematic errors on the $B^-\to\psi(2S)\pi^-$ branching fraction.
}
\begin{ruledtabular}
\begin{tabular}{lc}                      
 Source & Uncertainty (\%)
\\\hline
Uncertainty in yield & $3.5$ \\         
Tracking error & $5.0$         \\         
Lepton  identification & $4.2$             \\
Pion  identification & $1.6$ \\          
MC statistics &      $0.3$ \\          
 Number of $B\overline{B}$ pairs & $1.4$  \\         
Daughter branching fractions & $3.4$ \\
$\Delta M$ requirement &  $0.5$
\\\hline
Total & $8.4$         \\          
\end{tabular} 
\end{ruledtabular}
\end{table}

The $CP$-violating charge asymmetry $\mathcal{A}$ is defined as :
\begin{equation}
  \mathcal{A} = \frac{\mathcal{B}(B^{-}\to\psi(2S)\pi^{-}) -  \mathcal{B}(B^{+}\to\psi(2S)\pi^{+})}{ \mathcal{B}(B^{-}\to\psi(2S)\pi^{-}) + \mathcal{B}(B^{+}\to\psi(2S)\pi^{+})}.   
\end{equation}
    We  extract branching fractions for  $B^+$ and $B^-$ samples separately  using  the same procedure as described above, except the  polynomial background shape is fixed to that obtained in the previous fit to the total $B^{\pm}\to\psi(2S) \pi^{\pm}$ yield. We obtain 
\begin{equation}
  \mathcal{A} = 0.022 \pm 0.085 \pm 0.016
\end{equation}
with corresponding signal yields of $89\pm11$ and $93\pm11$ for $B^+$ and $B^-$, respectively. Here  the  systematic error  includes contributions from the  uncertainty on the  yield extraction method, charge asymmetry in the pion identification efficiency  as well as a possible  detector bias   and  a possible difference between $B^{+}$ and $B^{-}$  signal shape parameters.  Among all these uncertainties, the only non-negligible one is  a possible detector bias: it  is estimated to be $0.016$ from  a study of  $B^{-} \to J/\psi K^{-}$.

 We also measure the ratio of  $\mathcal{B}(B^{-}$~$\to$~$\psi(2S)$$\pi^{-}) $ and  $\mathcal{B}(B^{-}$~$\to$~$\psi(2S)$$K^{-}) $. In a  study using   large  $B$~$\to$~$(J/\psi,$ $\psi(2S))$$ X$  MC samples,  the amount of background for $B^{-}$~$\to$~$\psi(2S)$$(\ell^+ \ell^-)K^{-}$ is found to be   negligible. On the other hand, in the case of $B^{-} \to \psi(2S)(J/\psi \pi^+ \pi^-)K^{-}$, the decay mode  $B^{-}\to J/\psi K_1(1270)^{-}$ as well as $B^{-} \to $ $J/\psi \pi^+ \pi^- K^{-}$  appear as peaking backgrounds since  they have the same final state  as the signal.  The background estimation  is performed  using the $\Delta M$ data sideband defined as  $0.51$ GeV$/c^2 < \Delta M$$<0.57$ GeV$/c^2$ and $0.61$ GeV$/c^2 < \Delta M$$<0.67$ GeV$/c^2$. We obtain $7.2\pm2.7$ events  for  $B^{-}\to\psi(2S)(J/\psi(ee) \pi\pi)K^{-}$ and $9.4\pm2.7$ events  for  $B^{-}\to\psi(2S)(J/\psi(\mu\mu) \pi\pi)K^{-}$   when scaled to the signal region. These backgrounds are  subtracted from   the appropriate  signal yields. A simultaneous unbinned  extended maximum-likelihood fit gives a  branching fraction  of ($6.12\pm0.09$)$\times 10^{-4}$ and corresponding signal yield of $4720 \pm 69$, where  the error is  statistical only.  The resulting fit is shown in Fig.~3. The branching fraction is  in good agreement with our  previous measurement   \cite{abe1} and the  world-average \cite{PDG2006}.

\begin{figure}[h!]
\begin{center}
\includegraphics[scale=0.45]{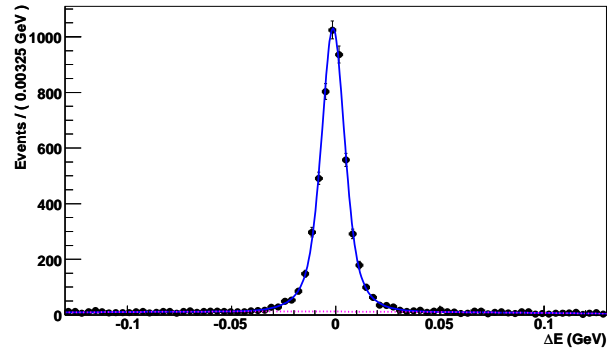}
\caption{\label{fig:qm/complexfunctions} $\Delta E$ distribution of the  $B^{-}\to\psi(2S)K^{-}$ candidates. The curves  show the overall fit  and the background component (magenta dotted for  combinatorial background). }
\end{center}
\end{figure}
We obtain the ratio of  $\mathcal{B}(B^{-}\to\psi(2S)\pi^{-}) $ 
and  $\mathcal{B}(B^{-}\to\psi(2S)K^{-}) $  to be  
 \begin{equation}
  \frac{\mathcal{B}(B^{-}\to\psi(2S)\pi^{-}) }{ \mathcal{B}(B^{-}\to\psi(2S)K^{-}) } =  (3.99\pm 0.36 \pm 0.17)\%,
\end{equation}
which is consistent with the expectations of  the factorization hypothesis.  Many sources of systematic errors  cancel in the ratio of the branching fractions.  Contributions to the systematic error come from the uncertainty in pion identification, signal extraction method, MC statistics  from $B^{-}\to\psi(2S)\pi^{-}~(K^{-})$ decay, kaon identification  and uncertainty on the background estimation from $B^{-}\to\psi(2S)K^{-}$ decay. The total uncertainty is  $4.2\%$. 

In summary,  we report the first  observation of the decay $B^{-}$ $\to$ $\psi(2S)$$\pi^{-}$  using $657 \times 10^{6} $ $B\overline{B}$ events.  The measured branching fraction is  $\mathcal{B}(B^{-}\to\psi(2S)\pi^{-})$ $=$ $(2.44\pm 0.22 \pm 0.20 )\times 10^{-5}$.  No significant direct $CP$-violating charge asymmetry is observed in $B^{-}\to\psi(2S)\pi^{-}$.  The ratio of branching fractions is $ (3.99 \pm 0.36\pm 0.17)\%$ which is consistent with the theoretical prediction based on factorization \cite{neubert}.

We thank the KEKB group for excellent operation of the accelerator, the KEK cryogenics group for efficient solenoid operations, and the KEK computer group and
the NII for valuable computing and SINET3 network support.  We acknowledge support from MEXT and JSPS (Japan); ARC and DEST (Australia); NSFC (China); DST (India); MOEHRD, KOSEF and KRF (Korea); KBN (Poland); MES and RFAAE (Russia); ARRS (Slovenia); SNSF (Switzerland); NSC and MOE (Taiwan); and DOE (USA).

%We thank the KEKB group for the excellent operation of the  accelerator, the KEK cryogenics group for the efficient  operation of the solenoid, and the KEK computer group and the National Institute of Informatics for valuable computing and Super-SINET network support. We acknowledge support from the Ministry of Education, Culture, Sports, Science, and Technology of Japan and the Japan Society for the Promotion of Science; the Australian Research Council and the Australian Department of Education, Science and Training; the National Natural Science Foundation of China under contract No.~10575109 and 10775142; the Department of Science and Technology of India; the BK21 program of the Ministry of Education of Korea, the CHEP SRC program and Basic Research program  (grant No.~R01-2005-000-10089-0) of the Korea Science and Engineering Foundation, and the Pure Basic Research Group program of the Korea Research Foundation; the Polish State Committee for Scientific Research; %-> remove for now: under contract No.~2P03B 01324; the Ministry of Education and Science of the Russian Federation and the Russian Federal Agency for Atomic Energy; the Slovenian Research Agency;  the Swiss National Science Foundation; the National Science Council and the Ministry of Education of Taiwan; and the U.S.\ Department of Energy.

%\newpage %Just because of unusual number of tables stacked at end


\begin{thebibliography}{99}
\bibitem{chargeconju}
Hereafter charge-conjugate modes are included throughout the paper 
unless stated otherwise.
\end{thebibliography}

\end{document}